\documentclass[aps,pre,showpacs,superscriptaddress,twocolumn]{revtex4-1}

\usepackage{amsfonts}
\usepackage{subfigure}
\usepackage{amsmath}
\usepackage{txfonts}
\usepackage{amssymb}
\usepackage{amsbsy} 
\usepackage{epsfig}
\usepackage{graphicx}
\usepackage{epstopdf}

\usepackage{mathdots} 

\def\be{\begin{equation}} \def\ee{\end{equation}}
\def\bea{\begin{eqnarray}} \def\eea{\end{eqnarray}}

\def\be{{\bf e}}

\begin{document}
\title{Detecting topological phases via survival probabilities of edge Majorana fermions}
\author{Yucheng Wang}
\thanks{wangyc@iphy.ac.cn}
\affiliation{Beijing National
Laboratory for Condensed Matter Physics, Institute of Physics,
Chinese Academy of Sciences, Beijing 100190, China}
\affiliation{School of Physical Sciences, University of Chinese Academy of Sciences, Beijing, 100049, China}
\begin{abstract}
We investigate the evolutions of edge Majorana fermions (MFs) and unveil that they can be used to characterize different topological phases and study the topological phase transitions. For some limiting cases of the evolution process for the Kitaev chain model and Su-Schrieffer-Heeger (SSH) model, we give analytical expressions of the survival probabilities (SPs) of the edge MFs. For a general case, we consider a dimerized Kitaev model as well as the Kitaev chain with disorder chemical potential and numerically calculate the SPs of two edge correlation MFs. Our results show that both of them equal to zero, one of them equals to zero at some times or neither of them equal to zero correspond to the SSH-like topological, topological superconductor and trivial phases respectively.
\end{abstract}
\pacs{05.30.Rt, 74.40.Gh, 03.65.Vf}
\maketitle
\section{Introduction}
The topological insulators and topological superconductors have attracted intensive studies in the last decade and a series of outstanding achievements have been achieved \cite{Zhang}, which have deepened our understanding of the basic physical properties of solid. Kitaev chain \cite{Kitaev} is a spinless p-wave superconductor system and it provides a promising candidate to explore Majorana fermions (MFs)\cite{Alicea,Gangadharaiah,Stoudenmire,Chan,Loss}, which fulfill non-Abelian statistics and are a potential application for the topological quantum computing \cite{Nayak,Quiroga}. Su-Schrieffer-Heeger (SSH) model \cite{ssh} is another famous one-dimensional (1D) model including the transitions between topological and trivial phases, which is regarded as a platform to study rich topological phenomena such as topological soliton excitations and topological edge states \cite{ssh2,Maki,Sarma}.
The topological property of a system can be characterized by calculating the topological indices in momentum space. The occurrence of topological phase transitions corresponds to the sudden change of topological indices. As the bulk-edge correspondence, one can analyze edge states in topological insulator or edge MFs in topological superconductors (TSCs) under open boundary conditions (OBC) to study a system's topological properties.

Dynamics of quantum systems have attracted a great deal of attentions \cite{Heyl,Heyl2} recently. The survival probability (SP) of an initial state is an important physical quantity in the dynamics research and it
characterizes the revival of this initial state later in time \cite{Jalabert,Gorin,Quan,Jafari}. This quantity also plays an important role in studying the extended-localization transition \cite{Geisel,Huckestein}. If the initial state is located in a lattice site, the SP of this state approaches to zero for a extended system but tends to a finite value of $O(1)$ for a localized system when the time $t\rightarrow \infty$ \cite{Santos}. For a topological system, although bulk states are extended, there exist zero-energy states that are localized at the edges of this system under OBC.
If the initial state is taken as the particle located in an edge, one can investigate it's SP after a long time evolution.

It is an interesting problem whether different topological phases can be distinguished by dynamics, or rather, by the survival probabilities (SPs) of edge states. To solve this question, we investigate a dimerized Kitaev model \cite{Nagaosa,Chen,Wang,Ezawa,Chitov} which includes the SSH-like topological, TSC and trivial phases. In this work, we introduce the SPs of two edge correlation MFs, each of them includes two SPs of edge MFs. We find that the two defined SPs can be used to distinguish different topological phases of the dimerized Kitaev model even added disorder chemical potential.

The paper is organized as follows. In section II, we introduce SPs of edge MFs and edge correlation MFs. We further give analytical expressions of the defined SPs for some limiting cases of the Kitaev chain model and SSH model. In section III, we numerically study the topological phase transitions of the general cases of the dimerized Kitaev model and the Kitaev chain model with disorder chemical potential by using the introduced SPs. A brief summary is given in section IV.

\section{Survival probability of Majorana fermions and topological phase transtions}
\subsection{Model Hamiltonian}
We consider a dimerized Kitaev superconductor chain \cite{Nagaosa,Chen,Wang,Ezawa,Chitov} with Hamiltonian
\bea
H &=& -J \sum_{j}^{L/2} [(1+\lambda)c^\dagger_{2j-1}c_{2j} + (1-\lambda)c^\dagger_{2j}c_{2j+1}+H.c.]\nonumber\\
  &-& \Delta \sum_{j}^{L/2}[(1+\lambda)c^{\dagger}_{2j-1}c^{\dagger}_{2j}+ (1-\lambda)c^\dagger_{2j}c^\dagger_{2j+1} +H.c.]\nonumber\\
  &+& \sum_{j}^{L/2}(\mu_{2j-1}c^\dagger_{2j-1}c_{2j-1}+\mu_{2j}c^\dagger_{2j}c_{2j}),
\label{ham-dimerized}
\eea
where $c_{j}^{\dagger}$ ($c_{j}$) is the fermionic creation (annihilation) operator, $J$ denotes the nearest-neighbor hopping strength and is taken as the unit of energy, i.e. $J=1$,  $\Delta$ is the superconducting pairing gap which is taken to be real here and $L$ is this system size. For the on-site chemical potential $\mu_j$, we will firstly discuss its site-independent case (all $\mu_j$ are set as $\mu$) and then study its randomized case in the last part of this work. We next take $|\lambda|\leq 1$ and OBC unless otherwise stated.
One can easily find that the Hamiltonian becomes the 1D Kitaev model \cite{Kitaev} when $\lambda=0$ and it reduces to the SSH model \cite{ssh} when $\Delta=0$ and $\mu=0$. Therefore this model should include three phases: SSH-like topological phase, TSC phase and topologically trivial phase \cite{Nagaosa,Chen,Wang,Ezawa,Chitov}.

we introduce the MF operators:
\begin{equation}
\gamma_{2j-1} = c_{j}+c^{\dagger}_{j}, \gamma_{2j} = \frac{1}{i}(c_{j}-c^{\dagger}_{j}),
\label{Majorana}
\end{equation}
which fulfill
\begin{equation}
\gamma_{j}^{\dagger}=\gamma_{j}, \left\{ \gamma_{j},\gamma_{l}\right\} =2\delta_{j l},
\end{equation}
By using the MF operators, the Hamiltonian (\ref{ham-dimerized}) can be written as
\begin{eqnarray}
&H=\frac{i}{2}&\sum_{j} [-(J+\Delta)(1+\lambda)\gamma_{4j-1}\gamma_{4j-2} \nonumber\\
&&-(J-\Delta)(1+\lambda)\gamma_{4j-3}\gamma_{4j} \nonumber\\
&&-(J+\Delta)(1-\lambda)\gamma_{4j+1}\gamma_{4j} \nonumber\\
&&-(J-\Delta)(1-\lambda)\gamma_{4j-1}\gamma_{4j+2}].
\label{ham-d2}
\end{eqnarray}
which is a $2L \times 2L$ matrix.

\subsection{SP of edge correlation MFs}
Given an initial state $|\Psi(0)\rangle$ at $t=0$, the evolution state at time $t$ can be written as
\begin{equation}
\begin{aligned}
|\Psi(t)\rangle=e^{-iHt}|\Psi(0)\rangle=\sum_{m}e^{-iE_mt}|\psi_m\rangle\langle\psi_m|\Psi(0)\rangle,
\label{final}
\end{aligned}
\end{equation}
where $|\psi_m\rangle$ is the $m-th$ eigenstate of the Hamiltonian $H$ and $E_m$ is the corresponding eigenvalue.  The SP of this initial state at the time $t$ can be defined as
 \begin{equation}
\begin{aligned}
 P(t)=|\langle\Psi(0)|\Psi(t)\rangle|^2=|\sum_{m}e^{-iE_mt}|\langle\psi_m|\Psi(0)\rangle|^2|^2,
\label{probability}
\end{aligned}
\end{equation}
which is related to the Loschmidt echo \cite{Jalabert,Gorin,Quan,Jafari,Yang}.

If the initial state is set as a MF locating at the first Majorana site of the Majorana chain with size $2L$, i.e., $|\Psi(0)\rangle=\gamma_{1}(0)|\Omega\rangle$, where $|\Omega\rangle$ is the Bogoliubov vacuum \cite{Bermudez}, its SP can be defined as $P_1(t)=|\left\langle\Omega| \gamma_{1}(0)\gamma_{1}(t)|\Omega\right\rangle|^2$ \cite{Rajak,Goldstein,explain} (see Ref.[31] for a detailed derivation). In the case of no causing ambiguity, we omit the $\Omega$. We can further define $P_{2L}(t)=|\left\langle \gamma_{2L}(0)\gamma_{2L}(t)\right\rangle|^2$, where the initial state is set as a MF locating at the $2L-th$ Majorana site. In a similar way, the SPs of the second and the $(2L-1)-th$ Majorana sites can be defined as $P_2(t)=|\left\langle \gamma_{2}(0)\gamma_{2}(t)\right\rangle|^2$ and $P_{2L-1}(t)=|\left\langle \gamma_{2L-1}(0)\gamma_{2L-1}(t)\right\rangle|^2$.

\begin{figure}
\includegraphics[height=70mm,width=80mm]{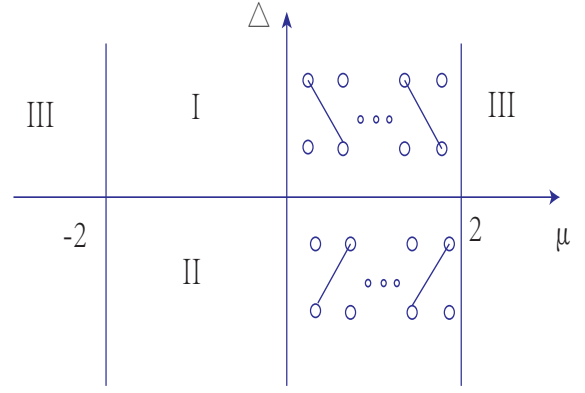}
\caption{\label{0}
  Phase diagram of the 1D Kitaev model ($\lambda=0$ for the Hamiltonian (\ref{ham-dimerized})). Phase $I$ and phase $II$ are the TSC phase and phase $III$ corresponds to the topologically trivial phase. The coupling cases of MFs in the regions $I$ and $II$ correspond to two limiting cases that $\Delta=J$ and $\Delta=-J$ with fixed $\mu=0$.}
\end{figure}
Because the MFs always appear in pairs, we consider the SP of edge correlation MFs \cite{Wang} with the initial state $\Psi^{1}(0)=\gamma_{1}\gamma_{2L}|\Omega\rangle$ after a long time evolution \cite{Goldstein} that
\begin{eqnarray}
G^{1}(t) & = & |\left\langle \gamma_{2L}(0)\gamma_{1}(0)\gamma_{1}(t)\gamma_{2L}(t)\right\rangle| \nonumber \\
 & = & |\left\langle \gamma_{1}(0)\gamma_{1}(t)\right\rangle||\left\langle \gamma_{2L}(0)\gamma_{2L}(t)\right\rangle|,
\end{eqnarray}
as our above definition, $G^{1}(t)=\sqrt{P_1(t)P_{2L}(t)}$. In a similar way, we can define the other SP of edge correlation MFs that $G^{2}(t)=\sqrt{P_2(t)P_{2L-1}(t)}$ with the initial state $\Psi^{2}(0)=\gamma_{2}\gamma_{2L-1}|\Omega\rangle$. Both $G^{1}$ and $G^{2}$ aren't equal to zero at any time $t$ means that there exists a Dirac fermion which robustly locate at the each end of this chain. If one of them is equal to zero at some times $t$ but the other one never equals to zero at any time, there exists a MF that robustly localize at the each end of this chain, which indicates that this system is TSC. If both of them simultaneously equal to zero at some times $t$, this system is topologically trivial. In order to obtain an intuitive understanding, we first consider some limiting cases of the Kitaev chain model as well as the SSH model.

\begin{figure}
\includegraphics[height=80mm,width=80mm]{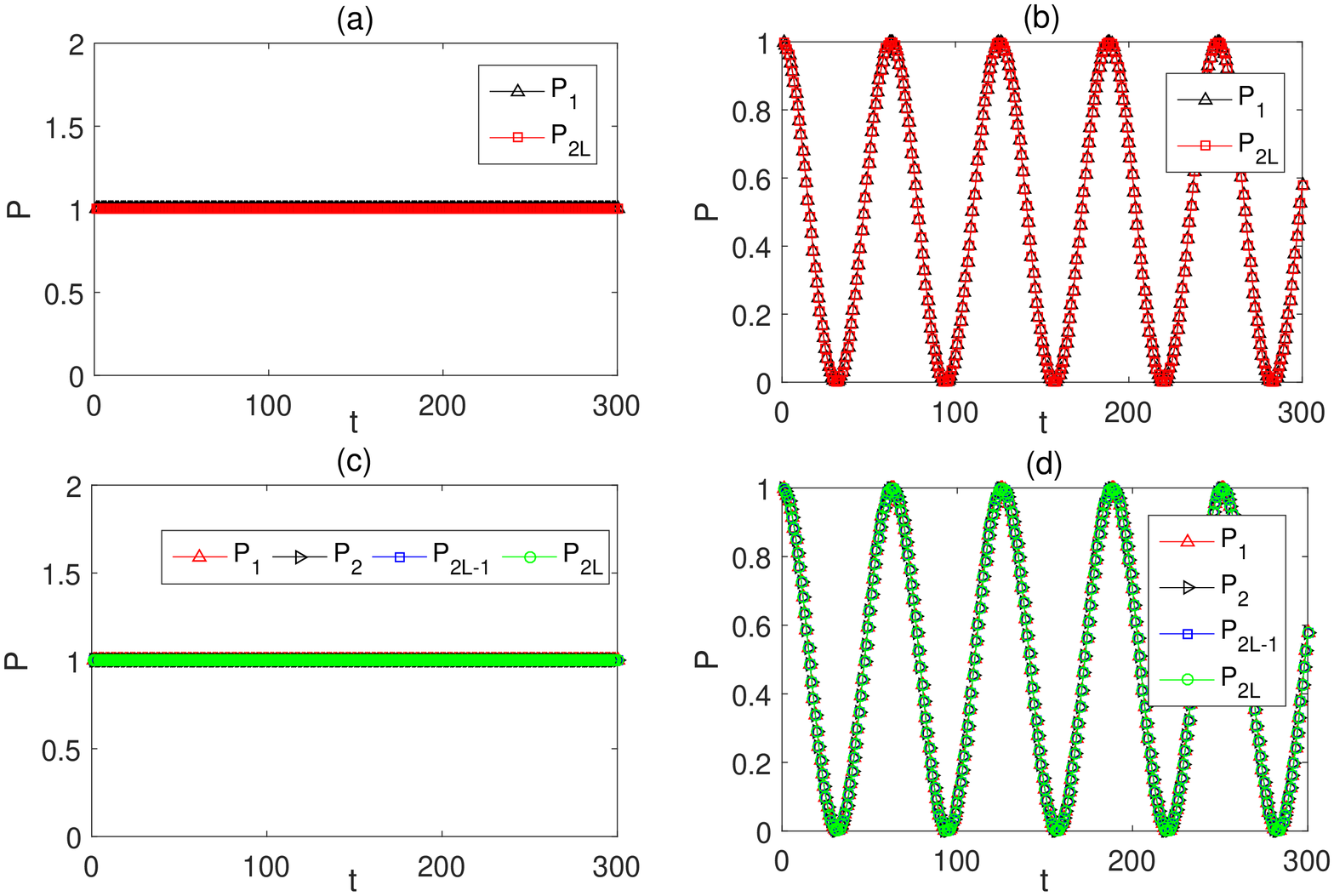}
\caption{\label{00}
  (Color online) The SP of the MFs $P_1$ and $P_{2L}$ for the 1D Kitaev model as a function of time $t$ for (a) $\Delta=-1$ and $\mu=0$, (b) $\Delta=J=0$ and $\mu=0.05$.
  $P_1$, $P_2$, $P_{2L-1}$ and $P_{2L}$ for the SSH model as a function of time $t$ for (c) $\lambda=-1$ and $J=0.025$, (d) $\lambda=1$ and $J=0.025$. The size of them are $L=600$.}
\end{figure}
\subsection{Kitaev chain model ($\lambda=0$)}
When $\lambda=0$, the Hamiltonian (\ref{ham-dimerized}) reduces to the 1D Kitaev model \cite{Kitaev}:
\bea
H &=& \sum_{j=1}^{L-1} (-J c^\dagger_{j}c_{j+1} - \Delta c^{\dagger}_{j}c^{\dagger}_{j+1} +H.c.)\nonumber\\
  &+& \sum_{j=1}^{L}\mu (c^\dagger_{j}c_{j}-\frac{1}{2}).
\label{ham-1}
\eea
The phase diagram of this system is showed in Fig.~\ref{0}, which is trivial when $|\frac{\mu}{2J}|>1$ and is TSC when $|\frac{\mu}{2J}|<1$.

By using the MF operators in Eq.(\ref{Majorana}), the Hamiltonian (\ref{ham-1}) can be written as
\bea
H &=& \frac{i}{2}\sum_{j}^{L-1} [(-J - \Delta) \gamma_{2j-1}\gamma_{2j+2} +(J-\Delta)\gamma_{2j}\gamma_{2j+1}]\nonumber\\
  &-& \frac{i}{2}\sum_{j}^{L}\mu \gamma_{2j-1}\gamma_{2j}.
\label{ham-2}
\eea

In the first case, we fix $\Delta=-J$ and $\mu=0$, then the Hamiltonian (\ref{ham-2}) becomes $H=iJ\gamma_{2j}\gamma_{2j+1}$. From Fig.~\ref{0}, this system is at TSC phase.
We can calculate the evolution of $\gamma_{1}$ and $\gamma_{2L}$,
\bea
\frac{d\gamma_1}{dt} = \frac{1}{i}[\gamma_1,H]=0,
\eea
and $\frac{d\gamma_{2L}}{dt}=0$. If there exists one MF located at the first or $2L-th$ Majorana site of this Majorana chain at $t=0$, the corresponding SP of this MF $P_{1}$ ($P_{2L}$) will be $1$ all the time, as showed in Fig.~\ref{00}(a), where we show the $P_{1}$ and $P_{2L}$ as a function of the time $t$ for this system with $\Delta=-J$ and $\mu=0$. One can see that $P_1(t)=P_{2L}(t)$ at any time, so we have $G^{1}(t)=P_1(t)=P_{2L}(t)$ as the above definition and $G^{1}(t)$ is always $1$ for this case. If MFs are located at the second and $(2L-1)-th$ Majorana sites of this Majorana chain at $t=0$, one can easily obtain that $G^{2}(t)$ oscillates with time and it will equal to zero at some times.
By contrast, we take $\Delta=J$ and $\mu=0$ and can easily verify that $G^{2}(t)$ is always $1$ at any time but $G^{1}$ equals to zero at some times. Our results are consistent with that this system is TSC under these parameters.

Next we consider $\Delta=J=0$, then this Hamiltonian becomes $H=-\frac{i\mu}{2}\sum_{j}^{L}\gamma_{2j-1}\gamma_{2j}$. We have
\begin{subequations}
\begin{eqnarray}
\frac{d\gamma_1}{dt} = \frac{1}{i}[\gamma_1,H]=-\mu\gamma_{2},\\
\frac{d\gamma_2}{dt} = \frac{1}{i}[\gamma_2,H]=\mu\gamma_{1},
\label{ta1}
\end{eqnarray}
\end{subequations}
so $\frac{d^{2}\gamma_1}{dt^2}=-\mu^2\gamma_{1}$, then it is easy to obtain $\gamma_{1}(t)=a\cos(\mu t)+b\sin(\mu t)$, where $a$ and $b$ are the undetermined operators. In a similar way, we obtain $\gamma_{2}(t)=c\cos(\mu t)+d\sin(\mu t)$, where $c$ and $d$ are the undetermined operators. If the initial state is set as $|\Psi(0)\rangle=\gamma_1(0)|\Omega\rangle$, we have $a=\gamma_1(0)$ and $c=0$. From Eq.(\ref{ta1}), we can obtain $d=\gamma_1(0)$ and $b=0$. Therefore, the SP of this MF $P_{1}$ will become $\cos^2(\mu t)$, which oscillates as the time period $T=\frac{\pi}{\mu}$ and it becomes zeros at the times $t=n\frac{\pi}{2\mu}$, where $n=1,2,3,\cdots$. Similarly, we can obtain $P_{2L}=\cos^2(\mu t)$ when the initial state is set as the MF located at the $2L-th$ Majorana site of the Majorana chain. Fig.~\ref{00}(b) displays $P_{1}$ and $P_{2L}$ versus $t$ for this model with $\Delta=J=0$ and $\mu=0.05$. It can be seen that $P_{1}$ and $P_{2L}$ oscillate in synchrony with the time period $T=\frac{\pi}{\mu}\approx 62.8$ and $P_{1}$, $P_{2L}$ and $G^{1}$ become zeros at the times $t=n\frac{T}{2}$, which is consistent with our analytical results. If the MFs located at the second and $(2L-1)-th$ Majorana sites of the Majorana chain is set as the initial state, one can easily verify that there exist same oscillations for $P_{2}$, $P_{2L-1}$, $G^{2}$ and they become zeros at the same times $t=n\frac{T}{2}$, which means that the message of the initial edge MFs completely disappear at these times and this system is topologically trivial.

\subsection{SSH model ($\Delta=0$ and $\mu=0$)}
When $\Delta=0$ and $\mu=0$, the Hamiltonian (\ref{ham-dimerized}) becomes the SSH model \cite{ssh}, which is written as:
\bea
H &=& -J \sum_{j=1}^{L/2} [(1+\lambda)c^\dagger_{2j-1}c_{2j} + H.c.]\nonumber\\
  &-& J\sum_{j=1}^{L/2-1}[(1-\lambda)c^\dagger_{2j}c_{2j+1}+H.c.]
\label{ham-ssh}
\eea
This system is topological when $\lambda<0$ and it is trivial when $\lambda>0$.

After introducing the Majorana operators as showed in Eq.(\ref{Majorana}), the Hamiltonian (\ref{ham-ssh}) becomes
\bea
H &=& \frac{-i J(1+\lambda)}{2}\sum_{j=1}^{L/2} (\gamma_{4j-3}\gamma_{4j} +\gamma_{4j-1}\gamma_{4j-2})\nonumber\\
  &-& \frac{i J(1-\lambda)}{2}\sum_{j=1}^{L/2-1} (\gamma_{4j-1}\gamma_{4j+2} +\gamma_{4j+1}\gamma_{4j}).
\label{ham-ssh2}
\eea
We then consider two limiting cases. In the first case, we fix $\lambda=-1$, then the Hamiltonian (\ref{ham-ssh2}) becomes $H=-iJ\sum_{j=1}^{L/2-1} (\gamma_{4j-1}\gamma_{4j+2} +\gamma_{4j+1}\gamma_{4j})$, which doesn't include $\gamma_{m}$, where $m=1,2,2L-1,2L$, so we have $[\gamma_{m}, H]=0$. Therefore,
$\frac{d\gamma_m}{dt} = \frac{1}{i}[\gamma_m,H]=0$, which means that if there exists one MF $\gamma_m$ located at the end of this chain at $t=0$, the SP of this MF will not decay with time. In Fig.~\ref{00}(c), we depict the $P_{m}$ as a function of the time $t$ for this system with $\lambda=-1$ and $J=0.025$ and we see that the SP of these MFs are always $1$. That is to say, both $G^{1}$ and $G^{2}$ are always $1$, which means that there exist two MFs (i.e., a Dirac fermion) which robustly locate at the each end of this system and it is a SSH-like topological phase.

The second limiting case is $\lambda=1$, then the Hamiltonian (\ref{ham-ssh2}) becomes $H = -iJ\sum_{j=1}^{L/2} (\gamma_{4j-3}\gamma_{4j} +\gamma_{4j-1}\gamma_{4j-2})$. It can be easily proved that $\gamma_m(t)=\gamma_{m}(0) cos(2Jt)$. The corresponding SP $P_{m}$ will become $\cos^2(2J t)$, which oscillates as the time period $T=\frac{\pi}{2J}$ and it equals to zero at the times $t=n\frac{\pi}{4J}$, where $n$ is a positive integer. Fig.~\ref{00}(d) shows that $P_{m}$ ($m=1,2,2L-1,2L$) versus $t$ with the parameters $J=0.025$ and $\lambda=1$. We see that all $P_{m}$ oscillate as the time period $T$ and they are equal to zero at the times $t=n\frac{T}{2}$. As the above definition, $G^{1}$ and $G^{2}$ simultaneously equal to zero at the times $t=n\frac{T}{2}$, which means that the initial edge MFs will completely disappear at these times and this system is topologically trivial.

Although $P_1(t)$ ($P_2(t)$) equals to $P_{2L}(t)$ ($P_{2L-1}(t)$) at any time for the above cases, there also exist some cases that $P_1(t)$ ($P_2(t)$) may not equal to $P_{2L}(t)$ ($P_{2L-1}(t)$), e.g., for the case that the system size $L$ is odd for the SSH model. Thus, in order to investigate the topological properties of a system, we need consider the four SPs $P_{1}$, $P_{2}$, $P_{2L-1}$ and $P_{2L}$ or just consider the two SPs that $G^{1}$ and $G^{2}$, which give the same results. For simplicity, we will just consider $G^{1}$ and $G^{2}$.

\begin{figure}
\includegraphics[height=80mm,width=80mm]{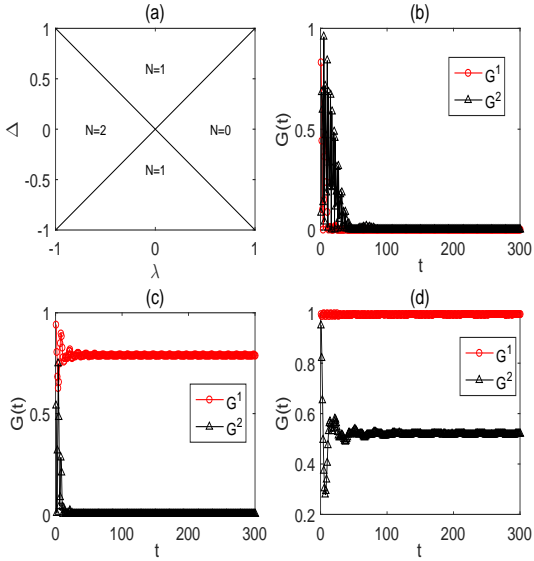}
\caption{\label{01}
  (Color online) (a) Topological phase diagram for the case in the absence of chemical potential. The vertical axis is $\Delta$ and the horizontal axis is $\lambda$. $N=2$, $N=1$ and $N=0$ denote the SSH-like topological phase, TSC phase and trivial phase respectively. The SPs of the edge correlation MFs $G^{1}$ and $G^{2}$ as a function of time $t$ for (b) $\lambda=0.7$, (c) $\lambda=0$ and (d) $\lambda=-0.7$. Here we fix the parameters $\Delta=0.5$ and $L=600$.}
\end{figure}

\begin{figure}
\includegraphics[height=70mm,width=80mm]{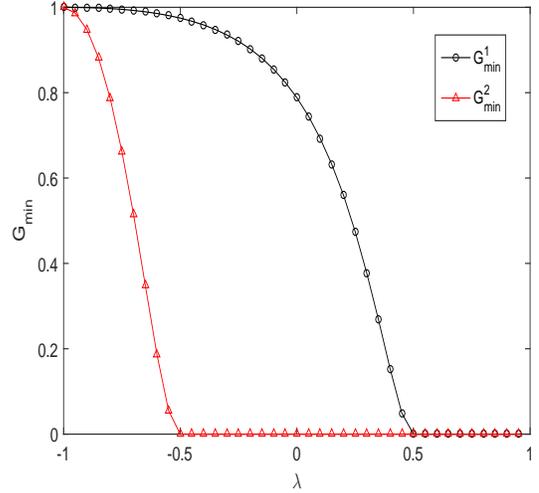}
\caption{\label{02}
(Color online) The minimum values of the SPs of the edge correlation MFs $G^{1}$ and $G^{2}$ for $t\in [100,300]$ as a function of $\lambda$ with fixed $\Delta=0.5$, $L=600$.}
\end{figure}

\section{Numerical study for general cases}
In this section, we investigate the general cases that the dimerized Kitaev model as well as the Kitaev chain model with disorder chemical potential. Although no topological invariants can be analytically defined for the latter case, we can still explore whether the presence or absence of the zeros of the SPs $G^{1}$ and $G^{2}$ at some times, which can be served as a characteristic signature of topological nontrivial or trivial phases.
\subsection{Topological phase transitions of the dimerized Kitaev model}
When the chemical potential is fixed as $\mu=0$ in the Hamiltonian (\ref{ham-dimerized}), the corresponding phase diagram of this system is presented in Fig.~\ref{01}(a), which can be obtained by calculating the topological numbers under periodic boundary conditions \cite{Nagaosa} or two edge correlation functions of MFs under OBC \cite{Wang}. The SPs of the edge correlation MFs $G^{1}$ and $G^{2}$ versus time $t$ for the topologically trivial phase, TSC and SSH-like topological phase are showed in Fig.~\ref{01}(b), (c) and (d) respectively. From these figures, we see that if the system is at the topologically trivial phase, both of the SPs can reach nearby zero after a finite time evolution, which means that the messages of the initial edge Majorana states completely disappear. If the system is at the TSC phase, one of the SPs can reach nearby zero but the other one approaches a nonzero constant after a finite time evolution. If the system is at the SSH-like topological phase, both of the SPs never approach zero after a long time evolution. Experimentally, the initial state that the MFs located in the first and $2L-th$ Majorana sites can be prepared by choosing the parameters that $\Delta=-J, \lambda=0$ and the initial state that the MFs located in the second and $(2L-1)-th$ Majorana sites can be prepared if we choose the parameters that $\Delta=J, \lambda=0$, then quenching the Hamiltonian to investigate the evolutions of these initial states \cite{Patel,Rajak,Sacramento,Sacramento2}.

From Fig.~\ref{01}, we see that the changes of $G^{1}$ and $G^{2}$ versus the time are oscillations during a short time and then almost become constants after a long time evolution, which has similarities with the dynamical evolution that discussing the localized-extended transition with the initial state located one lattice site \cite{Geisel,Huckestein}. One can easily find that $G^{1}$ and $G^{2}$ hardly change over time when $t>100$ from these figures. In Fig.~\ref{02}, we present the minimum values of $G^{1}$ and $G^{2}$ for $t\in [100, 300]$ as a function of $\lambda$ with fixed $\Delta=0.5$, $L=600$. Actually, these minimum values approximately equal to the average values of $G^{1}$ and $G^{2}$ for $t\in [100,300]$, since they hardly oscillate in that time. As was expected, both $G^{1}_{min}$ and $G^{2}_{min}$ don't equal to zero when $\lambda<-0.5$, this system is at the SSH-like topological phase, $G^{2}_{min}=0$ and $G^{1}_{min}\neq 0$ when $-0.5<\lambda<0.5$, this system is at TSC phase and both $G^{1}_{min}$ and $G^{2}_{min}$ equal to zero when $\lambda>0.5$, this system is at the topologically trivial phase. By using this method, we can also numerically determine the system's phase diagram as showed in Fig.~\ref{01} (a).

\begin{figure}
\includegraphics[height=120mm,width=80mm]{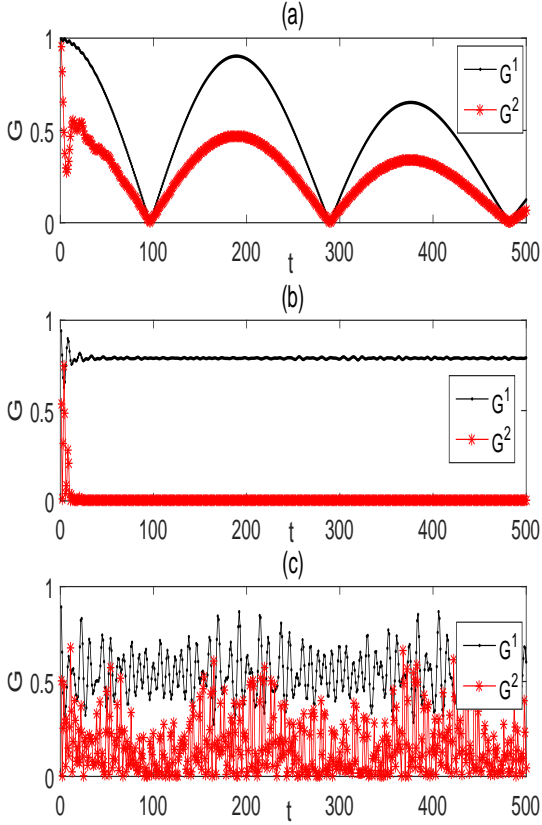}
\caption{\label{03}
(Color online) The SPs of the correlation MFs $G^{1}$ and $G^{2}$ as a function of the time $t$ with fixed $\Delta=0.5$, $L=600$ and (a) $\lambda=-0.7$, $w=0.1$, (b) $\lambda=0$, $w=0.1$, (c) $\lambda=0$, $w=3$.}
\end{figure}
\subsection{Kitaev chain model with disorder chemical potential}
We consider the uniformly distributed random potential $\mu_{j} \in [\frac{-w}{2},\frac{w}{2}]$ in the Hamiltonian (\ref{ham-dimerized}) to investigate the effect of disorder on this model. Fig.~\ref{03} (a) and (b) show that $G^{1}$ and $G^{2}$ as a function of time $t$ for this system with $w=0.1$ at the SSH-like topological phase and TSC phase respectively. We see that the SSH-like topological phase is sensitive but the TSC phase is robust against the disorder, this is because the disorder chemical potential breaks the sublattice symmetry but doesn't break the particle-hole symmetry of the superconductivity \cite{Nagaosa,Hegde}. Even the disorder strength $w$ is added to $3$, as showed in Fig.~\ref{03} (c), $G^{1}$ never approaches zero after a long time evolution, which means that the MFs are located at the two edges of this system and it is still TSC.

\begin{figure}
\includegraphics[height=70mm,width=80mm]{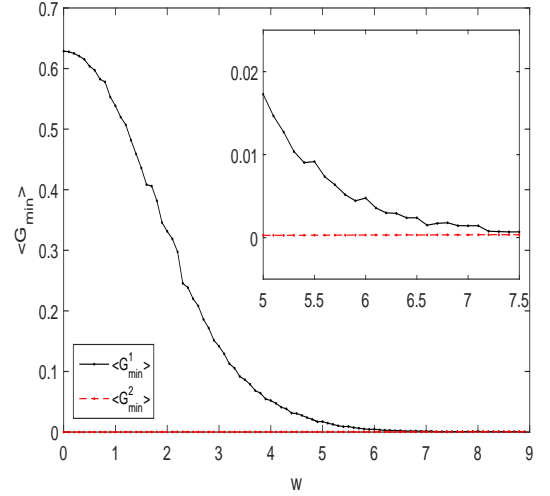}
\caption{\label{04}
(Color online) The averaged minimum values of $G^{1}$ and $G^{2}$ as a function of the disorder strength $w$ with fixed $\lambda=0$, $\Delta=0.5$, sample size=5000 and $L=400$. The inset shows the enlarged region from $w=5$ to $w=7.5$.}
\end{figure}

We further consider the topological phase transition of the Kitaev chain model induced by the disorder chemical potential, i.e., we set $\lambda=0$ for the Hamiltonian (\ref{ham-1}), which have been widely studied \cite{Tezuka,Lobos,Sen,Brouwer,Brouwer2,Sen2} and we now investigate it from dynamics. From Fig.~\ref{03} (c), $G^{1}$ and $G^{2}$ don't approach a constant after a long time evolution, so we take the minimum values of $G^{1}$ and $G^{2}$ during a long time. If both of them are nearby zero, this system is topologically trivial. If one of them is nearby zero but the other one doesn't approach zero at any time $t$, this system is TSC. To decrease the oscillation, we take some samples to obtain the averaged values of the minimum $G^{1}$ and $G^{2}$.  Fig.~\ref{04} show the averaged values of the minimum $G^{1}$ and $G^{2}$ in terms of the disorder strength $w$ under the fixed $\Delta=0.5$. We see that $\langle G^{1}_{min}\rangle \neq 0$ and $\langle G^{2}_{min}\rangle=0$ when the disorder strength is weak, which means that the system is at the TSC phase. Both $\langle G^{1}_{min}\rangle$ and $\langle G^{2}_{min}\rangle$ approximately equal to zero when $w$ is plenty big enough, which means that this system enters into a trivial phase. As disorder increases, a TSC-topologically trivial phase transition takes place. The transition point $w_{c}$ should satisfy that $\langle G^{1}_{min}\rangle > \langle G^{2}_{min}\rangle$ when $w<w_{c}$ and $\langle G^{1}_{min}\rangle \approx \langle G^{2}_{min}\rangle$ when $w_{c}+\delta w$, where $\delta w$ is a infinitely small quantity. From Fig.~\ref{04}, we see that the transition point is about $w_c\approx 7.3$, which is consistent with previous results \cite{Sen2}.

\section{Summary and discussions}
In summary, We have introduced the SPs of edge MFs and edge correlation MFs after a long time evolution to describe the topological phase transitions of a dimerized Kitaev model. When $\lambda=0$, this model reduces to the Kitaev model and we discussed two limiting cases $\Delta=-J$, $\mu=0$ and $\Delta=J=0$. We have obtained the analytical expressions with respect to  the SPs of edge MFs, which suggest that if there exist a serious of zero points at some times $t$, the system is trivial, otherwise it is TSC. When $\Delta=0$ and $\mu=0$, this model reduces to the SSH model and we discussed two limiting cases $\lambda=-1$ and $\lambda=1$. Our results show that all of them equal to zero at some time points $t$ for the trivial phase and all of them aren't equal to zero at any time corresponding to the SSH-like topological phase.
We further numerically investigated the two SPs of edge correlation MFs for a general dimerized Kitaev model and the Kitaev chain with disorder chemical potential. Our results show that if both of them aren't equal to zero at any time, the system is at a SSH-like topological phase, if one of them is equal to zero at some times but the other one isn't, the system is at the TSC phase and if both of them equal to zero at some times, this system is topologically trivial.

For a system in the presence of the disorder potential or interactions, it is difficult to define or calculate the topological invariant. In this paper, we provided an alternative solution to characterize different topological phases and topological phase transitions. Furthermore, it is very difficult to observe the topological invariant of a system in the experiment \cite{Xiongjun}. By contrast, we can detect the SP of a initial state in terms of the experiment. As was mentioned previously, the initial state can be prepared as that the MFs located in the first and $2L-th$ Majorana sites or the second and $(2L-1)-th$ Majorana sites by choosing some appropriate parameters \cite{Perfetto}. After the sudden variation of the system parameters according to different protocols, we can experimentally study the temporal evolution of the MFs to characterize different topological phases.

\begin{acknowledgments}
I thank Prof. S. Chen and C. Yang for helpful discussions.
\end{acknowledgments}



\begin{thebibliography}{25}
\bibitem{explain} $P_1(t)=|\left\langle\Omega| \gamma_{1}(0)\gamma_{1}(t)|\Omega\right\rangle|^2$= $|\left\langle\Omega| \gamma_{1}(0)e^{-iHt}\gamma_{1}(0)e^{iHt}|\Omega\right\rangle|^2$= $|\left\langle\Omega| \gamma_{1}(0)e^{-iHt}\gamma_{1}(0)e^{iE_0t}|\Omega\right\rangle|^2$= $|\left\langle\Omega| \gamma_{1}(0)e^{-iHt}\gamma_{1}(0)|\Omega\right\rangle|^2$= $|\langle\Psi(0)|e^{-iHt}|\Psi(0)\rangle|^2$= $|\langle\Psi(0)|\Psi(t)\rangle|^2$. $|\Omega\rangle$ is actually the ground state of this system and $E_0$ is the corresponding eigenvalue.
\end{thebibliography}
\end{document}